\documentstyle[a4,12pt,epsf]{article}
\renewcommand{\theequation}{\arabic{equation}}
\renewcommand{\thesection}{\arabic{section}}
\renewcommand{\thefootnote}{\fnsymbol{footnote}}
\newcommand{\bea}{\begin{eqnarray}}
\newcommand{\ena}{\end{eqnarray}}
\newcommand{\vs}[1]{\vspace{#1 mm}}

\newcommand{\z}{\omega}

\newcommand{\PL}[1]{Phys.\ Lett.\ {\bf #1}}

\newcommand{\NP}[1]{Nucl.\ Phys.\ {\bf #1}}
\newcommand{\PR}[1]{Phys.\ Rev.\ {\bf #1}}
\newcommand{\PRL}[1]{Phys.\ Rev.\ Lett.\ {\bf #1}}
\newcommand{\PTP}[1]{Prog.\ Theor.\ Phys.\ {\bf #1}}

\newcommand{\EPJ}[1]{Eur.\ Phys.\ J.\ {\bf #1}}

\newcommand{\boca}[1]{\mbox{\boldmath $#1$}}

\newcommand{\wt}[1]{\widetilde{#1}}

\newcommand{\gsim}{\stackrel{>}{_\sim}}

\begin{document}
\noindent

\begin{titlepage}
\setcounter{page}{0}
\begin{flushright}
July, 2000\\
OU-HET 353\\
hep-ph/0007066\\
\end{flushright}
\vs{2}
\begin{center}
{\Large{\bf
Quantum effects for the neutrino mixing matrix \\
in the democratic-type model}}\\
\vs{4}
{\large
Takahiro Miura\footnote{e-mail address:
miura@het.phys.sci.osaka-u.ac.jp},
Eiichi Takasugi\footnote{e-mail address:
takasugi@het.phys.sci.osaka-u.ac.jp}\\
\vs{2}
{\em Department of Physics,
Osaka University \\ Toyonaka, Osaka 560-0043, Japan} \\
\vs{4}
Masaki Yoshimura\footnote{e-mail address:
myv20012@se.ritsumei.ac.jp}\\
\vs{2}
{\em Department of Physics,
Ritsumeikan University \\ Kusatsu, Shiga 525-8577, Japan} }
\end{center}
\vs{5}
\centerline{{\bf Abstract}}
We investigate the quantum effects for
the democratic-type neutrino mass matrix given
at the right-handed neutrino mass scale $m_R$ in order to see
(i) whether $\theta_{23}=-\pi/4$ predicted by the model
is stable to explain the atmospheric neutrino anomaly,
(ii) how $\theta_{12}$ and $\theta_{13}$ behave, and
(iii) whether the predicted Dirac CP phase $\delta$ keeps maximal
size, at
the weak scale $m_Z$.
We find that,
for the (inversely) hierarchical mass spectrum with $m_1\sim m_2$,
$\theta_{23}$ and $\theta_{13}$ are stable,
while $\theta_{12}$ is not so, which leads to the possibility that
the solar neutrino mixing angle can become large at $m_Z$ even if
it is taken small at $m_R$.
We also show that $\delta$ keeps almost maximal
for the above mass spectrum,
and our model can give the large CP violation effect in the future
neutrino oscillation experiments if the solar neutrino puzzle
is explained by the large mixing angle MSW solution.

\end{titlepage}

\newpage
\renewcommand{\thefootnote}{\arabic{footnote}}
\setcounter{footnote}{0}

\section{Introduction}
Recent neutrino experiments have been strengthening the evidence of
the
neutrino mixing \cite{Mann:1999,Suzuki:1999}.
The study of the neutrino mixing opens a new phase for our deeper
understanding of neutrino physics.

Let us summarize the present experimental data on the neutrino mixing.
From the recent analysis of the atmospheric neutrino anomaly,
we have the following allowed regions of the mixing angle and
the mass squared difference as \cite{Mann:1999}
\bea
\sin^22\theta_{\rm atm}= 0.85 \sim 1\;,\qquad
\Delta m^2_{\rm
atm}=2\times10^{-3}\sim6\times10^{-3}\;(\mbox{eV}^2)\;.
\ena

The oscillation interpretation for the solar neutrino problem has
still
several parameter choices as \cite{Suzuki:1999,Suzuki_etal:2000}
\footnote{
In summary, we shall add a short remark for our results
with the latest report on {\it Neutrino 2000}
\cite{Suzuki_etal:2000}.}
\begin{itemize}
\item The large mixing angle (LMA) MSW solution
\bea
\sin^22\theta_{\rm LMA}= 0.5 \sim 1\;,\qquad
\Delta m^2_{\rm
LMA}=1\times10^{-5}\sim1\times10^{-4}\;(\mbox{eV}^2)\;,
\ena
\item The small mixing angle (SMA) MSW solution
\bea
\sin^22\theta_{\rm SMA}= 10^{-3} \sim 2\times10^{-2}\;,\qquad
\Delta m^2_{\rm SMA}=4\times10^{-6}\sim10^{-5}\;(\mbox{eV}^2)\;,
\ena
\item The vacuum oscillation (VO) solution
\bea
\sin^22\theta_{\rm VO}= 0.75 \sim 1\;,\qquad
\Delta m^2_{\rm VO}=10^{-11}\sim 10^{-10}\;(\mbox{eV}^2)\;.
\ena
\end{itemize}

In our earlier paper \cite{Fukuura_etal:2000,Miura_etal:2000-2},
we proposed a democratic-type mass matrix for left-handed
Majorana neutrinos. This model has quite special predictions,
$\theta_{23}=-\pi/4$ and $\delta=\pi/2$, where $\theta_{23}$
is the mixing angle between mass eigenstates $\nu_2$ and $\nu_3$,
and $\delta$ is the CP violation phase in the standard
parameterization of the mixing matrix \cite{Caso_etal:1998}.
The other mixing angles $\theta_{12}$ and $\theta_{13}$ are
left free (independent on neutrino masses).
The maximum value $\theta_{23}=-\pi/4$ is essential to explain
the large atmospheric neutrino mixing. Also the prediction,
$\delta=\pi/2$ is interesting under the situation where
the search for the model to predict the CP violation phase
is an urgent problem. We examined the underlying symmetry
for the democratic-type mass matrix and found that
this mass matrix is derived by imposing the $Z_3$ symmetry
for the model with two up-type doublet Higgs bosons
\cite{Fukuura_etal:2000,Miura_etal:2000-2}.

Here we want to consider the situation where the
democratic-type mass matrix is derived from the
dimension five effective interaction with one up-type
doublet Higgs boson based on $Z_3$ symmetry.
Let us define
\bea
\Psi_1&=&\frac1{\sqrt 3}(\ell_e+\z^2 \ell_\mu+\z \ell_\tau)\;,
\nonumber\\
\Psi_2&=&\frac1{\sqrt 3}(\ell_e+\z \ell_\mu+\z^2 \ell_\tau)\;,
\nonumber\\
\Psi_3&=&\frac1{\sqrt 3}(\ell_e+\ell_\mu+\ell_\tau) \;,
\ena
where $\ell_i$ is the left-handed lepton doublet defined by,
say,  $\ell_e=(\nu_{eL}, e_L)^T$, and
$\z={\rm exp}(i2\pi/3)$ which satisfies $\z^3=1$ and $1+\z+\z^2=0$.
The fields $\Psi_i$ behave as
irreducible representations of $Z_3$ symmetry under
the permutation of $\ell_e$, $\ell_\mu$ and $\ell_\tau$,
\bea
\Psi_1\to \z \Psi_1\;,\;\;
\Psi_2\to \z^2 \Psi_2 \;,\;\;   \Psi_3\to \Psi_3 \;.
\ena
If we introduce a Higgs doublet that behaves
$H_u \rightarrow \z^2 H_u$,
then we can construct the $Z_3$ invariant dimension five
effective Lagrangian as
\bea
{\cal L}_{0 }
=- (m^0_1+\tilde{m}_1)\overline{(\Psi_1)^C} \Psi_1
                        \frac{H_u H_u}{u^2}
   -2\tilde{m}_1\overline{(\Psi_2)^C} \Psi_3 \frac{H_u H_u}{u^2}
  \;,
\ena
where $u=<H_u>$.
Here we introduce two kinds of symmetry breaking terms,
${\cal L}_{1}$ and ${\cal L}_{2}$ as
\bea
{\cal L}_{1}
&=&- (m^0_3+\tilde{m}_3)\overline{(\Psi_3)^C} \Psi_3
                        \frac{H_u H_u}{u^2}
   -2\tilde{m}_3\overline{(\Psi_1)^C} \Psi_2 \frac{H_u H_u}{u^2}
  \;,\nonumber\\
{\cal L}_{2}
&=&- (m^0_2+\tilde{m}_2)\overline{(\Psi_2)^C} \Psi_2
                        \frac{H_u H_u}{u^2}
   -2\tilde{m}_2\overline{(\Psi_1)^C} \Psi_3 \frac{H_u H_u}{u^2}
  \;.
\ena
We remind that all types of $Z_3$ symmetry breaking terms are
in ${\cal L}_{1}$ and ${\cal L}_{2}$.
Now we consider ${\cal L}_{\rm eff}=
{\cal L}_{0}+{\cal L}_{1}+{\cal L}_{2}$. Then,
after the symmetry breaking $<H_u>=u$, we obtain the
democratic-type mass matrix under the assumption that
all coefficients $m_i^0$ and $\tilde m_i$ are real (see
Sec.2).

In this paper, we consider that the democratic-type mass
matrix (or the dimension five effective interaction)
is given at the right-handed neutrino mass scale,
$m_R$ and see the predictions at $m_Z$ by using
the renormalization group.
The renormalization group effect for the dimension five
effective interaction have been examined intensively[7 - 15].
Casas {\it et al.} have investigated some general features of the
quantum effects for the neutrino mixing matrix
independent of the specific mass matrix \cite{Casas_etal:1999}.
Haba {\it et al.} have studied how the mixing angles behave
by choosing the simple real mass matrix which explains the
experimental data at $m_Z$ \cite{Haba_etal:1999}.

We are interested in\\
(i) whether our predicted value $\theta_{23}=-\pi/4$ is stable,\\
(ii) how two angles $\theta_{12}$ and $\theta_{13}$ behave
against quantum corrections, and\\
(iii) whether the Dirac CP phase $\delta$ keeps maximal size at $m_Z$
as that predicted at $m_R$.

First point (i) is indispensable for the democratic-type model to
explain the atmospheric neutrino anomaly,
and second point (ii) is interesting in view of searching the
possibility
that
the small angle $\theta_{12}$ can be produced at $m_Z$ from the
large angle $\theta_{12}$ at $m_R$, or vice versa,
in addition to their stable solutions.
Third point (iii) is also important for future neutrino experiments
in order to get the signals of CP and T violation in the lepton
sector.
These problems depend largely on the neutrino mass spectrum.
In more restricted model with hierarchical mass spectrum,
we investigated the points (i) and (ii), and showed that
neutrino mixing angles are stable for the fully hierarchical mass
spectrum,
while the solar neutrino mixing angle is unstable
for the hierarchical case with $m_1\simeq m_2$,
where $m_1$ and $m_2$ are the first and second mass eigenvalues
\cite{Miura_etal:2000}.
For the general democratic-type model, in addition to the above
results,
we find the possibility that LMA and/or VO solutions
can be realized at $m_Z$ even if the solar neutrino mixing angle
is small at $m_R$ for the hierarchical case with $m_2/m_1-1<<1$.
Also in this case, Dirac CP phase is almost left maximal,
and we can expect that CP and T violation effect can be detected
in the future long baseline neutrino experiment such as neutrino
factories
\cite{NF}.

This paper is organized as follows.
In section 2, we briefly review the democratic-type neutrino mass
matrix
and its predictions.
In section 3, we show the quantum corrections for the neutrino mass
matrix.
In section 4, we calculate the $\tan\beta$ dependence of
neutrino masses, mixing angles and Dirac CP phase.
In section 5, we give the results of numerical calculation
and compare with the analytical estimation.
Section 6 is devoted to the summary.

\section{Democratic-type mass matrix}

Throughout of this paper,
we assume that the mass matrix of charged leptons is diagonal.
After symmetry breaking, we obtain
the democratic-type neutrino  mass matrix from Eqs.(7),(8)
in the flavor eigenstate basis as[4, 5]
\bea
M_\nu(m_R)=\sum_{j=1}^3 (m_j^0 S_j + \wt{m}_j T_j)\;,
\ena
where
\bea
S_1&=&\frac{1}{3}\pmatrix{1& \z^2& \z \cr \z^2& \z& 1 \cr \z& 1&
\z^2 }\;,\quad
S_2=\frac{1}{3}\pmatrix{1& \z& \z^2 \cr \z& \z^2& 1 \cr \z^2& 1&
\z }\;,\quad
S_3=\frac{1}{3}\pmatrix{1& 1& 1\cr 1& 1& 1 \cr 1& 1& 1 }\;,\nonumber\\
T_1&=&\pmatrix{1& 0& 0 \cr 0& \z& 0 \cr 0& 0& \z^2 }\;,\quad
T_2=\pmatrix{1& 0& 0 \cr 0& \z^2& 0 \cr 0& 0& \z }\;,\quad
T_3=\pmatrix{1& 0& 0 \cr 0& 1& 0 \cr 0& 0& 1 }\;.
\ena
Six quantities $m_j^0$ and $\wt{m}_j$ ($j=1,2,3$) called mass
parameters
are taken to be real, and the form in Eq.(9) is assumed to
be generated from the effective dimension-five operators at $m_R$.

$M_\nu(m_R)$ can be transformed into real symmetric matrix
$\overline{M}_\nu(m_R)=V_{\rm Tri}^TM_\nu(m_R)V_{\rm Tri}$,
where $V_{\rm Tri}$ is the following tri-maximal mixing matrix as
\bea
V_{\rm Tri}=\frac{1}{\sqrt{3}}
\pmatrix{1& 1& 1\cr \z& \z^2& 1\cr \z^2& \z& 1}\;.
\ena
Then, the neutrino mixing matrix $U$ (MNS matrix
\cite{Maki_etal:1962})
which diagonalizes $M_\nu(m_R)$ as
$U^T M_\nu(m_R) U=D_\nu\equiv{\rm diag}(m_1, m_2, m_3)$ is expressed
as
$U=V_{\rm Tri}O$, where $O$ is the orthogonal matrix which
diagonalizes
$\overline{M}_\nu(m_R)$, and $m_i$ ($i=1,2,3$) are neutrino mass
eigenvalues.
This expression leads to the condition that the mixing
matrix $U$ should satisfy $U_{\mu i} = U_{\tau i}^\ast$ ($i=1,2,3$),
which do not depend on real mass parameters.
From these conditions on the mixing matrix, we find that
$c_{23}^2=s_{23}^2=1/2$ and $\cos\delta =0$
by using the standard parameterization advocated in
\cite{Caso_etal:1998}.
Here $c_{ij}=\cos\theta_{ij}$, $s_{ij}=\sin\theta_{ij}$ with
the mixing angle $\theta_{ij}$ between mass eigenstates $\nu_i$ and
$\nu_j$,
and $\delta$ is the Dirac CP phase.
Thus, we can obtain the following expression of $U$ as
\bea
U
=\pmatrix{1& 0& 0\cr 0& e^{i\rho}& 0\cr 0& 0& e^{-i\rho}}
\pmatrix{c_{12}c_{13}& s_{12}c_{13}& -is_{13}\cr
-\frac{s_{12}-ic_{12}s_{13}}{\sqrt{2}}&
\frac{c_{12}+is_{12}s_{13}}{\sqrt{2}}& -\frac{c_{13}}{\sqrt{2}}\cr
-\frac{s_{12}+ic_{12}s_{13}}{\sqrt{2}}&
\frac{c_{12}-is_{12}s_{13}}{\sqrt{2}}& \frac{c_{13}}{\sqrt{2}}}
\pmatrix{1& 0& 0\cr 0& 1& 0\cr 0& 0& i}\;,
\ena
where we have taken $s_{23}=-c_{23}=-1/\sqrt{2}$ and $\delta=\pi/2$.
The quantity $\rho$ is a redundant phase which can be absorbed into
charged leptons by the phase redefinition.
${\rm Diag}(1,1,i)$ is the Majorana phase matrix, which shows no
CP violation intrinsic to Majorana neutrinos.
Indeed, the phase $i$ relates to CP signs of neutrino masses
in addition to their relative sign assignments.
In this model, therefore, six real mass parameters
($m_i^0$, $\wt{m}_i$ ($i=1,2,3$)) are changed into three
neutrino masses ($m_1$, $m_2$, $m_3$),
two mixing angles ($\theta_{12}$, $\theta_{13}$),
and one unphysical phase ($\rho$).

Since three neutrino mass eigenvalues are free parameters,
we adopt the following mass squared differences as
\bea
\Delta m^2_{\rm atm}\equiv |\Delta m^2_{32}|\sim|\Delta
m^2_{31}|\;,\qquad
|\Delta m^2_{\rm 12}| << \Delta m^2_{\rm atm}\;,
\ena
where $\Delta m^2_{ij}\equiv m_i^2-m_j^2$.
Under the assignment of Eq.(13),
we consider the following mass spectrum as
\bea
\begin{array}{lll}
\mbox{Hierarchical case}&:& m_1\simeq m_2<<m_3\;,\\
\mbox{Inversely hierarchical case}&:& m_1\simeq m_2>>m_3\;,
\end{array}
\ena
and we assume all mass eigenvalues are positive
\footnote{
In this spectrum,
the behavior of the mixing matrix hardly depends on the sign of $m_3$.
So we can generally take $m_2>0$.
Then, the behavior of the mixing matrix
depends on the relative sign of $m_1$ and $m_2$
as well as their absolute sizes.
For $m_1<0$, mixing angles are stable while CP violation phase is
unstable.
See Ref. \cite{Miura_etal:2000}.
}.
Of course, there are another mass spectrums which satisfies Eq.(13);
fully hierarchical case ($m_1<<m_2<<m_3$) and
nearly degenerate case ($m_1\simeq m_2\simeq m_3$)
\cite{Altarelli_etal:1998}.
However,
for the former case, quantum corrections hardly change the
structure of the neutrino mass matrix, and hence all physical
quantities
are stable.
For the latter case,
neutrino mixing is highly sensitive to the input values at $m_R$,
and it is laborious to obtain their analytical expressions.
It is also noted
that we do not adopt that $|\Delta m^2_{12}|$ is the
mass squared difference for the solar neutrino mixing.
This is because mass eigenvalues $m_i$ ($i=1,2,3$) are those
given at $m_R$,
and as we shall see later,
$|\Delta m^2_{12}|$ varies while $|\Delta m^2_{32}|$ is almost stable
against quantum corrections, when the mass spectrum is given as
Eq.(14).

\section{Quantum corrections}

The neutrino mass matrix, and hence the neutrino mixing matrix,
may vary by quantum corrections
\cite{Chankowski_etal:1993,Babu_etal:1993}.
In the minimal supersymmetric standard model (MSSM)
with dimension-five operators which give Majorana masses to
left-handed neutrinos, quantum corrections appear in the relation
of the Majorana mass matrices between at $m_R$ and at $m_Z$
as \cite{Haba_etal:1999}
\bea
M_\nu(m_Z)=
\pmatrix{\frac{1}{\sqrt{I_e}}& 0& 0\cr 0& \frac{1}{\sqrt{I_\mu}}& 0\cr
0& 0&
\frac{1}{\sqrt{I_\tau}} }
M_\nu(m_R)
\pmatrix{\frac{1}{\sqrt{I_e}}& 0& 0\cr 0& \frac{1}{\sqrt{I_\mu}}& 0\cr
0& 0&
\frac{1}{\sqrt{I_\tau}} }\;,
\ena
where $I_i$ ($i=e,\mu,\tau$) are defined as
\bea
I_i={\rm exp}
\left(
\frac{1}{8\pi^2}\int_{\ln(m_R)}^{\ln(m_Z)}y_i^2dt
\right)\;.
\ena
Here $y_i$ are Yukawa couplings of charged leptons
in the mass eigenstate,
$t=\ln\mu$ with the renormalization point $\mu$,
and overall renormalization effect has been absorbed into
$M_\nu(m_R)$.

According to the discussion in \cite{Haba_etal:1999},
the approximation of $\sqrt{I_j/I_\tau}\sim1/\sqrt{I_\tau}$
($j=e,\mu$) is
held with good accuracy in the region of $2<\tan\beta<60$.
Here $\tan\beta=\langle\phi_u\rangle/\langle\phi_d\rangle$,
in which $\phi_u$ and $\phi_d$ are two Higgs doublets in the MSSM.
By using this approximation, $M_\nu(m_Z)$ reduces to the
following simpler form as
\bea
M_\nu(m_Z)
=\pmatrix{1& 0& 0\cr 0& 1& 0\cr 0& 0& \frac{1}{\sqrt{I_\tau}}}
M_\nu(m_R)\pmatrix{1& 0& 0\cr 0& 1& 0\cr 0& 0&
\frac{1}{\sqrt{I_\tau}}}
=M_\nu(m_R)-\epsilon M_1 + O(\epsilon^2)\;,
\ena
where
\bea
M_1=\pmatrix{
0& 0& (M_\nu(m_R))_{13}\cr
0& 0& (M_\nu(m_R))_{23}\cr
(M_\nu(m_R))_{13}& (M_\nu(m_R))_{23}& 2(M_\nu(m_R))_{33}}\;,
\ena
and $\epsilon$ is defined as
\bea
\epsilon=1-\frac{1}{\sqrt{I_\tau}}
=1-\left(\frac{m_Z}{m_R}\right)^{\frac{1}{8\pi^2}(1+\tan^2\beta)
(m_\tau/v)^2}>0\;,
\ena
with the mass of $\tau$ lepton, $m_\tau$, and
$v^2=\langle\phi_u\rangle^2+
\langle\phi_d\rangle^2$.
In the second equality in Eq.(19), we have neglected the running
effect
of $y_\tau$.
In order to estimate the value of $\epsilon$,
we consider the right-handed neutrino mass scale $m_R$ as $10^{13}$
GeV.
Then, with $m_Z=91.187$ GeV, $m_\tau=1.777$ GeV and $v=245.4$ GeV, we
find
\bea
8\times 10^{-5} < \epsilon < 6\times 10^{-2}\qquad
(\mbox{for}\quad 2< \tan\beta < 60)\;.
\ena
Therefore, we neglect the $O(\epsilon^2)$ terms in Eq.(17)
when obtaining the analytical expression
of the mixing angles and neutrino masses at $m_Z$.

By using Eq.(17), we can obtain the Majorana neutrino mass matrix
at $m_Z$.
This mass matrix depends on $\tan\beta$ via $\epsilon$ as well as
neutrino masses, mixing angles and CP phases given at $m_R$.
We take $\theta_{23}$ and CP phases at $m_R$ as those
predicted by the democratic-type model.
We also assume that neutrino masses at $m_R$ is given as Eqs.(13) and
(14).
Then we can investigate the $\tan\beta$ dependence of the neutrino
mixing matrix at $m_Z$,
which includes two mixing angles, $\theta_{12}$, $\theta_{13}$,
and neutrino masses at $m_R$.
As we will show in the next section,
$\theta_{13}$ hardly depends on quantum corrections,
so we can take the most stringent constraint on $\theta_{13}$,
$s_{13}=0.16$, from the CHOOZ data \cite{Apollonio_etal:1999}.

We also check the analytical estimation by numerical calculation,
which will be shown in section 5.
For any $\tan\beta$, we calculate the mass matrix $M_\nu(m_Z)$
numerically
by using Mathematica, and find the unitary matrix $\hat{U}$ which
diagonalizes $M_\nu(m_Z)$ as $\hat{U}^TM_\nu(m_Z)\hat{U}$.
Hereafter, we denote physical quantities at $m_Z$ as
$\hat\theta$, $\hat{m}_1$ and so on.
$\tan\beta$ dependence of the mixing angle $\hat\theta$
are shown by using the following expression as
\bea
\sin^22\hat\theta_{13}
&=&4|\hat{U}_{e3}|^2(1-|\hat{U}_{e3}|^2)\;,\nonumber\\
\sin^22\hat\theta_{23}
&=&4\frac{|\hat{U}_{\mu3}|^2}{1-|\hat{U}_{e3}|^2}
\left(1-\frac{|\hat{U}_{\mu3}|^2}{1-|\hat{U}_{e3}|^2}\right)\;,\nonumber\\
\sin^22\hat\theta_{12}
&=&4\frac{|\hat{U}_{e2}|^2}{1-|\hat{U}_{e3}|^2}
\left(1-\frac{|\hat{U}_{e2}|^2}{1-|\hat{U}_{e3}|^2}\right)\;.
\ena

\section{Quantum effects for physical quantities}
In this section, we show the $\tan\beta$ dependence of neutrino
masses,
mixing angles and Dirac CP phase at $m_Z$.

\subsection{neutrino masses and mixing angles}

By transforming $M_\nu(m_Z)$ by $U$ in Eq.(12),
we obtain the mass matrix $\wt{M}_\nu(m_Z)$
by keeping $\epsilon$ up to the first order as
\bea
\wt{M}_\nu(m_Z)
&=&U^T M_\nu(m_Z) U\nonumber\\
&=&\pmatrix{
(1-\epsilon|p|^2)m_1 &
\frac{1}{2}\epsilon(m_1pq^\ast+m_2p^\ast q) &
i\epsilon P\cr
\frac{1}{2}\epsilon(m_1pq^\ast+m_2p^\ast q) &
(1-\epsilon|q|^2)m_2 &
-i\epsilon Q\cr
i\epsilon P & -i\epsilon Q & (1-\epsilon c_{13}^2)m_3}\;,
\ena
where
\bea
p&\equiv& s_{12}-ic_{12}s_{13}\;,\quad
q\equiv c_{12}+is_{12}s_{13}\;,\nonumber\\
P&\equiv& \frac{1}{2}c_{13}(m_1p-m_3p^\ast)\;,\quad
Q\equiv \frac{1}{2}c_{13}(m_2q-m_3q^\ast)\;.
\ena

Let us define the submatrices as
\bea
\boca{\mu}&=&\pmatrix{
(1-\epsilon|p|^2)m_1 &
\frac{1}{2}\epsilon(m_1pq^\ast+m_2p^\ast q) \cr
\frac{1}{2}\epsilon(m_1pq^\ast+m_2p^\ast q) &
(1-\epsilon|q|^2)m_2}\;,\quad
\boca{m}=\pmatrix{i\epsilon P\cr -i\epsilon Q}\;,\nonumber\\
M&=&(1-\epsilon c_{13}^2)m_3\;.
\ena
Then, $\boca{m}$ is much smaller than either $M$
in the hierarchical case
or $\boca{\mu}$ in the inversely hierarchical case.
So we can block diagonalize $\wt{M}_\nu$ by
using seesaw expansion in \cite{Doi_etal:1988} as
\bea
U_{\rm seesaw}^T \wt{M}_\nu U_{\rm seesaw}
\simeq \pmatrix{\boca{\mu}& \boca{0}\cr \boca{0}& M}\;,
\ena
where
\bea
U_{\rm seesaw}\simeq
\pmatrix{\boca{1}_2& iS\cr iS^\dagger& 1}\;,
\ena
with a 2 by 2 unit matrix $\boca{1}_2$, and
\bea
iS\simeq
\left\{
\begin{array}{lc}
(M^{-1}\boca{m}^T)^\dagger
\simeq M^{-1}\pmatrix{-i\epsilon P^\ast\cr i\epsilon Q^\ast}
&\mbox{(hierarchical case)}\;,\\
-\boca{\mu}^{-1}\boca{m}
\simeq \mu^{-1}\pmatrix{-i\epsilon P m_2\cr  i\epsilon Q m_1}
&\mbox{(inversely hierarchical case)}\;,\\
\end{array}
\right.
\ena
where $\mu\equiv {\rm det}\boca{\mu}$.
Here we have neglected the normalization factor of $U_{\rm seesaw}$
since it is nearly unity.
By keeping $\epsilon$ up to the first order,
$U_{\rm seesaw}$ is simply rewritten as
\bea
U_{\rm seesaw}\simeq
\pmatrix{1& 0& \pm\frac{i}{2}\epsilon c_{13}p\cr
0& 1& \mp\frac{i}{2}\epsilon c_{13}q\cr
\pm\frac{i}{2}\epsilon c_{13}p^\ast&
\mp\frac{i}{2} \epsilon c_{13}q^\ast& 1}\;.
\ena
Here, the upper (lower) sign is for the hierarchical
(inversely hierarchical) case,
where we have neglected $O(m_{1,2}/m_3)$ ($O(m_3/m_{1,2})$) terms .

Now, in order to obtain the neutrino mixing matrix $\hat{U}$,
we only have to diagonalize the submatrix $\boca{\mu}$ in Eq.(25).
Let us define the small parameter
\bea
\xi=1-\frac{m_2}{m_1}\;.
\ena
By keeping $\epsilon$ and $\xi$ up to the first order,
$\boca{\mu}$ is rewritten as
\bea
\boca{\mu}
\simeq\pmatrix{
1-\epsilon|p|^2 & \frac{1}{2}\epsilon \sin2\theta_{12}c_{13}^2\cr
 \frac{1}{2}\epsilon \sin2\theta_{12}c_{13}^2& 1-\epsilon|q|^2-\xi
}m_1\;.
\ena
It is easy to diagonalize $\boca{\mu}$ since it is a 2 by 2 real
symmetric matrix.
By diagonalizing $\boca{\mu}$ as
$\wt{U}_{12}^T\boca{\mu}\wt{U}_{12}$
in which
\bea
\wt{U}_{12}=\pmatrix{
\cos\wt\theta& \sin\wt\theta\cr -\sin\wt\theta& \cos\wt\theta}=
\pmatrix{\wt{c}& \wt{s}\cr -\wt{s}& \wt{c}}\;,
\ena
we get
\bea
\tan2\wt\theta\simeq
-\frac{\epsilon\sin2\theta_{12}c_{13}^2}
{\epsilon\cos2\theta_{12}c_{13}^2+\xi}\;.
\ena

Then, the neutrino mixing matrix $\hat{U}$ is given as
\bea
\hat{U}\simeq UU_{\rm seesaw}\wt{U}\;,
\ena
and each element is written as follows
\bea
\hat{U}_{e1}&\simeq&
c_{13}\left[c'_{12}\left(1\mp\frac{1}{2}\epsilon s_{13}^2\right)
\pm\frac{i}{2}\epsilon s'_{12}s_{13}\right]\;,\nonumber\\
\hat{U}_{e2}&\simeq&
c_{13}\left[s'_{12}\left(1\mp\frac{1}{2}\epsilon s_{13}^2\right)
\mp\frac{i}{2}\epsilon c'_{12}s_{13}\right]\;,\nonumber\\
\hat{U}_{e3}&\simeq&
s_{13}\left(1\pm\frac{1}{2}\epsilon c_{13}^2\right)\;,\nonumber\\
\hat{U}_{\mu1}&\simeq&
-\frac{1}{\sqrt{2}}
\left[s'_{12}\left(1\mp\frac{1}{2}\epsilon c_{13}^2\right)
-ic'_{12}s_{13}\left(1\pm\frac{1}{2}\epsilon c_{13}^2\right)
\right]\;,\nonumber\\
\hat{U}_{\mu2}&\simeq&
\frac{1}{\sqrt{2}}
\left[c'_{12}\left(1\mp\frac{1}{2}\epsilon c_{13}^2\right)
+is'_{12}s_{13}\left(1\pm\frac{1}{2}\epsilon c_{13}^2\right)
\right]\;,\nonumber\\
\hat{U}_{\mu3}&\simeq&
-\frac{i}{\sqrt{2}}c_{13}
\left(1\pm\frac{1}{2}\epsilon c_{13}^2\right)\;,\nonumber\\
\hat{U}_{\tau1}&\simeq&
-\frac{1}{\sqrt{2}}(s'_{12}+ic'_{12}s_{13})
\left(1\pm\frac{1}{2}\epsilon c_{13}^2\right)\;,\nonumber\\
\hat{U}_{\tau2}&\simeq&
\frac{1}{\sqrt{2}}(c'_{12}-is'_{12}s_{13})
\left(1\pm\frac{1}{2}\epsilon c_{13}^2\right)\;,\nonumber\\
\hat{U}_{\tau3}&\simeq&
\frac{i}{\sqrt{2}}
\left[1\pm\frac{1}{2}\epsilon(1+s_{13}^2)\right]\;,
\ena
where
\bea
c'_{12}\equiv \cos\theta'_{12}
=\cos{(\theta_{12}+\wt{\theta})}\;,\quad
s'_{12}\equiv \sin\theta'_{12}
=\sin{(\theta_{12}+\wt{\theta})}\;,
\ena
and we have neglected an unphysical phase.

Thus, we obtain the mixing angle at $m_Z$ as
\bea
\sin^22\hat\theta_{13}&\simeq&
\sin^22\theta_{13}
\left(
1\pm\epsilon \cos2\theta_{13}
\right)\;,\nonumber\\
\sin^22\hat\theta_{23}&\simeq&
1+O(\epsilon^2)\;,\nonumber\\
\sin^22\hat\theta_{12}&\simeq&
\sin^22\theta'_{12}\left(1+O(\epsilon^2)\right)\nonumber\\
&=&\frac{(\xi\sin2\theta_{12})^2}
{(\epsilon c_{13}^2+\xi\cos2\theta_{12})^2+(\xi\sin2\theta_{12})^2}
\left(1+O(\epsilon^2)\right)\;.
\ena
Therefore, we may say that
there are no extra mixings between $\nu_{1,2}$ and $\nu_3$
by the renormalization group equation,
where $\nu_i$ is the mass eigenstate at $m_R$.
That is, mixing angles $\theta_{13}$ and $\theta_{23}=-\pi/4$ are
essentially
stable against quantum corrections.
Also the mixing angle $\hat\theta_{12}$ is almost equal to
$\theta'_{12}=\theta_{12}+\wt{\theta}$, and hence,
we may roughly obtain $\hat{U}$ by changing from $\theta_{12}$ to
$\theta'_{12}$ in $U$.
In other words, the behavior of the mixing angles hardly depends on
the contribution from seesaw expansion.

Next, we estimate the neutrino mass eigenvalues $\hat{m}_i$
($i=1,2,3$)
at $m_Z$.
They are given as
\bea
\hat{m}_1&\simeq&\frac{m_1}{2}
\left[
2-\epsilon(1+s^2_{13})-\xi
+{\rm sign}(\xi)\sqrt{(\xi+\epsilon \cos2\theta_{12} c_{13}^2)^2+
(\epsilon \sin2\theta_{12} c_{13}^2)^2}\right]\;,\nonumber\\
\hat{m}_2&\simeq&\frac{m_1}{2}
\left[
2-\epsilon(1+s^2_{13})-\xi
-{\rm sign}(\xi)\sqrt{(\xi+\epsilon \cos2\theta_{12} c_{13}^2)^2+
(\epsilon \sin2\theta_{12} c_{13}^2)^2}\right]\;,\nonumber\\
\hat{m}_3&\simeq&(1-\epsilon c_{13}^2)m_3\;,
\ena
where sign$(\xi)$ is $+1(-1)$ for $\xi>0(\xi<0)$.
Then, the mass squared difference $|\Delta \hat{m}_{12}^2|$ is given
as
\bea
|\Delta \hat{m}^2_{12}|
\simeq2m_1^2\sqrt{
(\xi+\epsilon \cos2\theta_{12} c_{13}^2)^2+
(\epsilon \sin2\theta_{12} c_{13}^2)^2
}\;.
\ena

As one can understand by looking at $\sin^22\hat\theta_{12}$ in
Eq.(36)
and at Eq.(38),
$\tan\beta$ dependences of $\sin^22\hat\theta_{12}$
and $|\Delta \hat{m}_{12}^2|$ depend on the sign of $\xi$.
Therefore, we consider the following two cases :

\vskip 5mm

\noindent
Case (a) : $\xi>0$ $(m_1>m_2)$

In this case, the first term of the denominator of
$\sin^22\hat\theta_{12}$
increases monotonously as $\tan\beta$ grows
\footnote{
Here we set $0\leq\theta_{12}\leq\pi/4$
since we consider that electron neutrinos mainly consist of $\nu_1$
if the solar neutrino mixing is small at $m_R$.
When $\theta_{12}=\pi/4$, i.e., solar mixing angle is exactly maximal
at the
$m_R$ scale, the dependences of $\sin^22\hat\theta_{12}$
and $\Delta \hat{m}^2_{12}$ do not depend on the sign of $\xi$.
}.
Thus, $\sin^22\hat\theta_{12}$ becomes smaller
as the quantum corrections become larger.
Hence, one may expect that SMA solution at $m_Z$ is realized
from the large mixing angle at $m_R$.
However, such undertaking is in vain.
Let us explain the reason briefly.
If one would try to find the
parameter regions of $\Delta m^2_{12}$ and $\epsilon$
to produce SMA solution,
one should at least set
$|\Delta \hat{m}_{12}^2|\simeq \Delta m^2_{\rm SMA}$
and
$\sin^22\hat\theta_{12}\simeq\sin^22\theta_{\rm SMA}$.
By simplifying $c_{13}^2=1$ from the CHOOZ bound,
one could obtain
\bea
|\Delta m^2_{12}|&\simeq&\Delta m^2_{\rm SMA}
\left|\frac{\sin2\theta_{\rm
SMA}}{\sin2\theta_{12}}\right|\;,\nonumber\\
\epsilon&\simeq&\frac{\Delta m^2_{\rm SMA}}{2m_1^2}
|\cos2\theta_{\rm SMA}|
\left(1-\cot2\theta_{12}\tan2\theta_{\rm SMA}\right)\;.
\ena
The first relation in Eq.(39) shows that $|\Delta m^2_{12}|$ is needed
to be
about 1/10 times as small as $\Delta m^2_{\rm SMA}$
when $\theta_{12}$ is large at $m_R$.
For example, we can obtain
$|\Delta m^2_{12}|\simeq 10^{-7}\sim10^{-6}\;{\rm eV}^2$
when $\sin2\theta_{12}=\sqrt{8/9}$
\footnote{This angle has been predicted in a
restricted democratic-type model \cite{Miura_etal:2000-2}.}.
In this case, however, the quantity $\Delta m^2_{12}$ is negative at
$m_Z$,
and as a result the matter effect cannot take place.



\vskip 5mm

\noindent
Case (b) : $\xi<0$ $(m_1<m_2)$

In this case, the first term of the denominator
of $\sin^22\hat\theta_{12}$ in Eq.(36) can become 0 when
\bea
\epsilon \simeq 
|\xi|\cos2\theta_{12}
=\frac{|\Delta m_{12}^2|}{2m_1^2}\cos2\theta_{12}\;,
\ena
and at the same time $\sin^22\hat\theta_{12}$ becomes maximal.
Here we have used $c_{13}^2=1$ for simplicity.
Hence we can expect that the large mixing angle such as LMA and VO
solutions
at $m_Z$ can be realized even from the small mixing angle
at $m_R$.
By substituting Eq.(40) into Eq.(38), we obtain the initial mass
splitting
as
\bea
|\Delta m^2_{12}|\simeq\frac{\Delta
\hat{m}^2_{12}}{\sin2\theta_{12}}\;.
\ena
Thus, the mass splitting at $m_R$ is
about 10 times as large as that at $m_Z$
when $\theta_{12}$ is the small mixing angle.
By setting $\sin2\theta_{12}=0.1$ such as preferred by SMA solution,
for example, we obtain
$|\Delta m^2_{12}|\simeq 1\times 10^{-4} \sim 1\times 10^{-3}\;{\rm
eV}^2$
for LMA and
$|\Delta m^2_{12}|\simeq 10^{-10} \sim 10^{-9}\;{\rm eV}^2$
for VO solutions.

In the inversely hierarchical case with
$\Delta m^2_{\rm atm}=3.5\times10^{-3}\;{\rm eV}^2$ and $s_{13}=0.16$,
we obtain $\epsilon \simeq 0.01\sim 0.1$ for LMA,
and $\epsilon \simeq 10^{-8}\sim 10^{-7}$ for VO solutions,
from Eq.(40).
Thus, LMA solution can be generated for $\tan\beta \gsim 30$,
while $\tan\beta$ is too tiny to produce VO solution for the
realistic $\tan\beta$ region.

On the contrary,
in the hierarchical case, $m_1^2$ is restricted to about
$2\times 10^{-3}\;{\rm eV}^2$
for LMA solution since $|\Delta m^2_{12}|$ can become as large as
$1\times 10^{-3}\;{\rm eV}^2$,
which leads to the results that LMA solution can be realized for the
large $\tan\beta$ region, i.e., $\tan\beta\gsim 40$.
However, we may say
$m_1^2\simeq 10^{-8}\sim 10^{-3}\;{\rm eV}^2$ for VO solution,
and VO solution can be realized for wide $\tan\beta$ region.

\subsection{CP violation phase}

Now, let us take our attention to the CP phase.
From Eq.(34), we get the Jarlskog parameter $\hat{J}$ as
\bea
|\hat{J}|
=|\mbox{Im}(\hat{U}_{e1}\hat{U}_{\mu2}
\hat{U}_{e2}^\ast \hat{U}_{\mu1}^\ast)|
\simeq\frac{1}{4}s_{13}c_{13}^2\sin2\theta'_{12}
\left[1\pm\frac{1}{2}\epsilon(1-3s_{13}^2)\right]\;.
\ena
The deviation from unity in the bracket means the contribution from
seesaw expansion.
This shows that the $\tan\beta$ dependence of the Jarlskog parameter
is almost same as that of $\sin2\theta'_{12}$.
That is, $\hat{J}$ is damping as $\tan\beta$ grows for $\xi>0$,
while it has a peak for $\xi<0$.

Dirac CP phase $\hat\delta$ is also given as
\bea
|\sin\hat\delta|
&=&\frac{|\mbox{Im}(\hat{U}_{e1}\hat{U}_{\mu2}
\hat{U}_{e2}^\ast \hat{U}_{\mu1}^\ast)|}
{|\hat{U}_{e1}||\hat{U}_{e2}||\hat{U}_{e3}|
|\hat{U}_{\mu3}||\hat{U}_{\tau3}|}(1-|\hat{U}_{e3}|^2)\nonumber\\
&\simeq&\sqrt{\frac{\sin^22\theta'_{12}}
{\sin^22\theta'_{12}+(\epsilon s_{13})^2}}\nonumber\\
&\simeq&
\sqrt{
\frac{(\xi\sin2\theta_{12})^2}
{(\xi\sin2\theta_{12})^2\left[1+(\epsilon s_{13})^2\right]+
(\epsilon s_{13})^2(\epsilon c_{13}^2+\xi\cos2\theta_{12})^2}
}\;.
\ena
Without the contribution from seesaw expansion,
we could not look at the corrections of the denominator of the middle
in Eq.(43), $(\epsilon s_{13})^2$,
and $|\sin\hat\delta|$ would not depend on the quantum correction.
In other words, we may say that Dirac CP phase is stable
for the small $\tan\beta$ region whatever
$\xi$ and $\theta_{12}$ are given at $m_R$.
However, $(\epsilon s_{13})^2$ have a possibility of becoming
comparable
to the $\sin^22\theta'_{12}$ in the large $\tan\beta$ region,
and we should take this contribution into account
to compare with the numerical evaluation.

\section{Numerical check}

We calculated the numerical evaluation of neutrino masses and
the mixing matrix at $m_Z$ to compare with the
analytical estimation shown in the previous section.
To simplify the analysis, we evaluated in the inversely hierarchical
case,
where the mass spectrum at $m_R$ is easy to be determined
except for $|\Delta m^2_{12}|$.
Figs. 1 and 2 are the results of the $\tan\beta$ dependence of
mixing angles, Dirac CP phase and mass squared differences at $m_Z$.

\vskip 5mm

\noindent
(1) An example of the case (a)

In this case, $\sin^22\hat\theta_{12}$ at $m_Z$ becomes
small for $\tan\beta>4$ as we can see from Eq.(36).
The numerical analysis is performed to see the $\tan\beta$
dependence for $\sin^22\hat\theta_{12}$ and other quantities
in detail and the result is shown in Fig. 1.
As we can see, $\sin^22\hat\theta_{12}$ can become as small as
0.01 at around $\tan\beta\sim8$, though
$\sin^22 \theta_{12}=8/9$ at $m_R$. The mass squared
difference $|\Delta m^2_{12}|$ at $m_Z$ become large.
For larger values of $\tan\beta$, $|\sin\hat\delta|$ becomes
small, though $|\sin \delta|=1$ at $m_R$. Other quantities,
$\hat\theta_{13}$, $\hat\theta_{23}$ and
$|\Delta \hat{m}_{23}^2|$ do not change much.

From Fig.1, one might think that this instability can be
used to realize SMA solution. Unfortunately, this
is not true, because $(m_2^2-m_1^2)\cos 2\hat\theta_{12}<0$
at $m_Z$ so that the MSW mechanism does not work.

\vskip 5mm

\noindent
(2) An example of the case (b)

In Fig. 2, we take as input values
$m_1=\sqrt{\Delta m^2_{\rm atm}}$,
$m_2=\sqrt{\Delta m^2_{\rm atm}+|\Delta m^2_{12}|}$,
$m_3=0$,
$\Delta m^2_{\rm atm}=3.4\times10^{-3}$ (eV$^2$),
$|\Delta m^2_{12}|=1\times 10^{-4}$ (eV$^2$),
$\sin2\theta_{12}=0.1$, $s_{13}=0.16$
with the predicted values of $\theta_{23}=-\pi/4$
and $\delta=\pi/2$ at $m_R$.
$\hat\theta_{13}$, $\hat\theta_{23}$ and
$|\Delta \hat{m}_{23}^2|$ are hardly dependent on the quantum
corrections
similarly as in the case (a).
$\sin^22\hat\theta_{12}$ has a peak around $\tan\beta\sim 30$.
From Eq.(40), we obtain the value of $\epsilon$ at the peak as
$1.5\times10^{-2}$,
which corresponds to $\tan\beta\sim30$,
which is consistent with the figure.
$|\Delta \hat{m}^2_{12}|$ is decreasing till $\tan\beta\sim30$ and
then increasing, which is caused by the negative $\xi$ in Eq.(38).
$|\sin\hat\delta|$ is almost maximal in the wide $\tan\beta$ region,
which is due to larger $\Delta m^2_{12}$ than that in the previous
example, and the effect of quantum corrections disappears.
Thus, the selection of those parameter values above
gives an example of generating
LMA solution with maximal CP phase at $m_Z$.

\section{Summary}

We investigated the quantum effects for the
democratic-type neutrino mass matrix
with the (inversely) hierarchical mass spectrum.
We assumed that this mass matrix is generated
by dimension-five operators added in the MSSM at $m_R$,
and considered the mass matrix at $m_Z$ by using the
renormalization group.
We summarize our results as follows :
\begin{itemize}
\item
$\theta_{23}=-\pi/4$, $\theta_{13}$ and $|\Delta m^2_{23}|$
are almost stable against quantum corrections.
\item
Dirac CP phase $\delta=\pi/2$ is almost stable
unless the input mass splitting $\xi=(m_1-m_2)/m_1$ at $m_R$
is too small.
\item The behavior of the mixing angle $\theta_{12}$
and $\Delta m^2_{12}$ is divided into
two cases :
\begin{itemize}
\item Case (a) : $m_1>m_2$ case\\
In this case $\sin^22\theta_{12}$ is damping dependent on
the size of the mass difference $\xi$.
From this nature, we expected the possibility of
obtaining SMA solution at $m_Z$
even if the solar neutrino mixing angle is large at $m_R$.
Though the mixing angle can become as small as the region
preferred by SMA solution,
$\Delta m^2_{12}\cos 2\hat\theta_{12}$ at $m_Z$ is negative and
the matter effect cannot occur in this case.
\item Case (b) :  $m_1<m_2$ case\\
In this case  $\sin^22\theta_{12}$ has a peak which is
dependent on the size of $\xi$.
From this nature, we showed the possibility of
obtaining LMA and/or VO solutions at $m_Z$
even if the solar neutrino mixing angle is small at $m_R$.
For the inversely hierarchical case, we can
obtain LMA solution in the region of $\tan\beta\gsim 30$.
For the hierarchical case, we can obtain either LMA solution
in the region of $\tan\beta\gsim 40$ or VO solution
in the wide $\tan\beta$ range.
In order for this phenomena to occur, $\Delta m^2_{12}$ at $m_R$
should be taken about 10 times as large as the experimental data.
\end{itemize}
\end{itemize}
\vskip 5mm

From the above results,
our democratic-type model can give the nearly maximal
mixing angle for the atmospheric neutrino anomaly at $m_Z$,
provided the free parameter $\theta_{13}$ is taken to be
as small as that preferred by the CHOOZ data.

As to the solar neutrino problem,
the latest report from Super-Kamiokande shows that
SMA and VO solutions are disfavored by comparing the
day/night spectrum with the results of the flux global analysis
\cite{Suzuki_etal:2000}.
If we consider this new data seriously,
putting too strong degeneracy on $m_1$ and $m_2$, i.e., $\xi<<1$, at
$m_R$
is limited in the small $\tan\beta$ region,
otherwise the solar neutrino mixing angle would become small at $m_Z$
no matter how large it could be taken at $m_R$.


The result that $\delta=\pi/2$ predicted by the democratic-type model
is almost kept maximal is the best situation to
search the CP and T violation phenomenon in the
near future projects like neutrino factories.
Thus, if the solar neutrino problem will be solved by LMA solution,
our model will be checked by looking at the
signals of CP and T violation by the neutrino oscillation experiments
in the next century.

\vskip 1cm
{\LARGE \bf Acknowledgment}\\
This work is supported in part by
the Japanese Grant-in-Aid for Scientific Research of
Ministry of Education, Science, Sports and Culture,
No.12047218.

\newpage
\setcounter{section}{0}
\renewcommand{\thesection}{\Alph{section}}
\renewcommand{\theequation}{\thesection .\arabic{equation}}
\newcommand{\apsc}[1]{\stepcounter{section}\noindent
\setcounter{equation}{0}{\Large{\bf{Appendix\,\thesection\,:\,{#1}}}}}


\begin{figure}[pht]
\epsfxsize=16cm
\centerline{\epsfbox{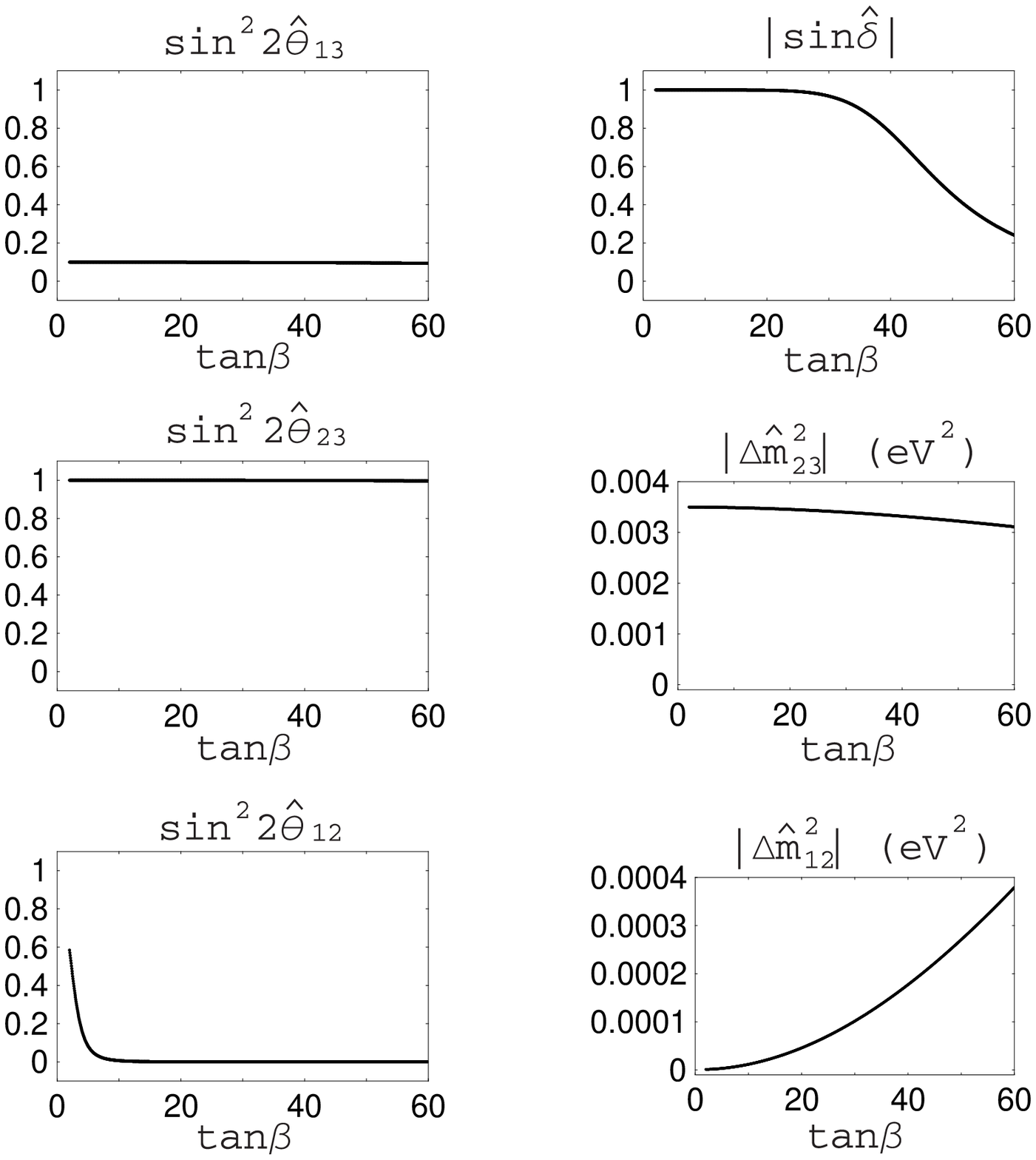}}
\caption{$\tan\beta$ dependence of neutrino mixing angles, Dirac CP
phase
and mass squared differences
for $m_1>m_2>>m_3$ at $m_R$.
As initial values at $m_R$, we took
$(m_1,m_2,m_3)=(5.9169\times 10^{-2},5.9161\times 10^{-2},0)$eV,
$\sin2\theta_{12}=\sqrt{8/9}$, $s_{13}=0.16$. We reminded that
$\theta_{23}=-\pi/4$ and $\delta=\pi/2$, which are the
predictions of our model.}
\end{figure}

\begin{figure}[pht]
\epsfxsize=16cm
\centerline{\epsfbox{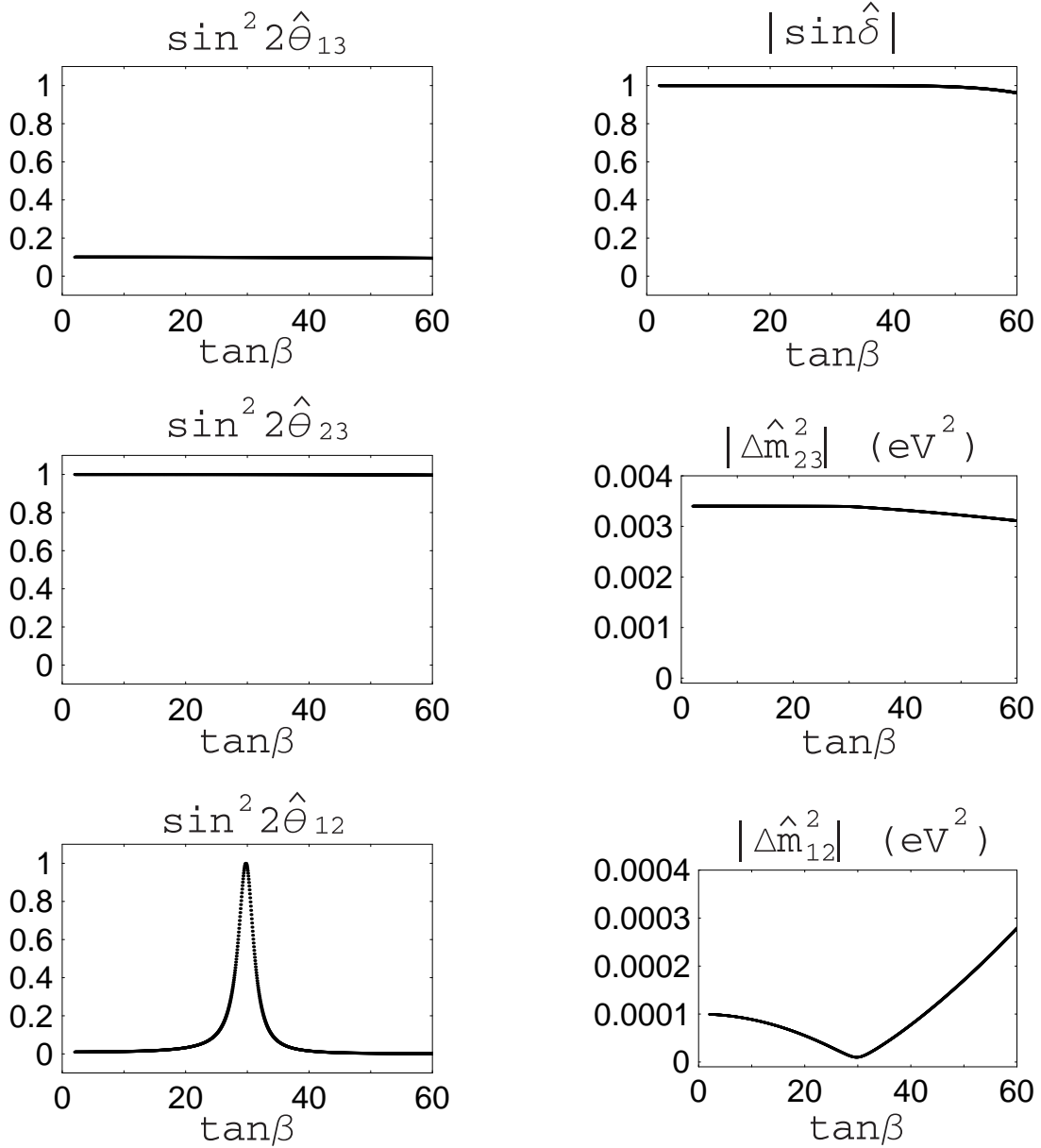}}
\caption{$\tan\beta$ dependence of neutrino mixing angles, Dirac CP
phase
and mass squared differences at $m_Z$ for $m_3<<m_1<m_2$.
As initial values at $m_R$, we took
$(m_1,m_2,m_3)=(5.831\times 10^{-2},5.916\times 10^{-2},0)$eV,
$\sin2\theta_{12}=\sqrt{8/9}$, $s_{13}=0.16$ with the model's
predictions $\theta_{23}=-\pi/4$ and $\delta=\pi/2$.}
\end{figure}

\end{document}